**Title**

# Vortex Ferroelectric Domains, Large-loop Weak Ferromagnetic Domains, and Their Decoupling in Hexagonal (Lu, Sc)FeO₃

*Kai Du, Bin Gao, Yazhong Wang, Xianghan Xu, Jaewook Kim, Rongwei Hu, Fei-Ting Huang and Sang-Wook Cheong\**

Dr. K. Du, Dr. B. Gao, Y. Wang, X. Xu, Dr. J. W. Kim, Dr. R. Hu, Dr. F. Huang and Prof. S.-W. Cheong*

Rutgers Center for Emergent Materials and Department of Physics and Astronomy, Rutgers University, Piscataway, New Jersey 08854, USA

* To whom the correspondence should be addressed. (E-mail: sangc@physics.rutgers.edu)

**Abstract**

The direct domain coupling of spontaneous ferroelectric polarization and net magnetic moment can result in giant magnetoelectric (ME) coupling, which is essential to achieve mutual control and practical applications of multiferroics. Recently, the possible bulk domain coupling, the mutual control of ferroelectricity (FE) and weak ferromagnetism (WFM) have been theoretically predicted in hexagonal $LuFeO_3$. Here, we report the first successful growth of highly-cleavable Sc-stabilized hexagonal $Lu_{0.6}Sc_{0.4}FeO_3$ (*h*-LSFO) single crystals, as well as the first visualization of their intrinsic cloverleaf pattern of vortex FE domains and large-loop WFM domains. The vortex FE domains are on the order of 0.1-1 μm in size. On the other hand, the loop WFM domains are ~100 μm in size, and there exists no interlocking of FE and WFM domain walls. These strongly manifest the decoupling between FE and WFM in *h*-LSFO. The domain decoupling can be explained as the consequence of the structure-mediated coupling between polarization and dominant in-plane antiferromagnetic spins according to the theoretical prediction, which reveals intriguing interplays between FE, WFM, and antiferromagnetic orders in *h*-LSFO. Our results also indicate that the magnetic topological charge tends to be identical to the structural topological charge. This could provide new insights into the induction of direct coupling between magnetism and ferroelectricity mediated by structural distortions, which will be useful for the future applications of multiferroics.

## Introduction

Multiferroic materials, in which two or multiple ferroic orders coexist, have drawn a great deal of attentions due to their fundamental importance and potentials for the next generation devices [1–3]. However, in most multiferroics, the magnetoelectric (ME) coupling strength is quite weak, especially in linear-ME materials [4], which limits their practical applications. Therefore, multiferroics with the direct domain coupling between spontaneous magnetization (*M*) and polarization (*P*) are highly sought after. The direct domain coupling can result in giant ME coupling, enabling the mutual control in the sense that flipping one of *M* or *P* can induce the flipping of the other. This direct domain coupling effect has been partially achieved, for example, at the hetero-interfaces of bilayer films. *M* in one film layer can be flipped by flipping *P* in the other film layer by electrical fields, however not the other way around [5]. The similar situation has also been discovered in single-phase $Dy_{0.7}Tb_{0.3}FeO_3$ and $Dy_{0.75}Gd_{0.25}FeO_3$ where only the flipping of *M* by electrical fields is reported [6]. Other examples in single-phase bulk systems are very limited. $CoCr_2O_4$ [7,8] and $Mn_2GeO_4$ [9] appear to be the only two single-phase materials that contain spontaneous *M* and magnetism-induced *P*, which can be flipped by the magnetic field. However, their *P* magnitude is tiny for any applications, and flipping *P* (thus, *M*) with electrical fields cannot be achieved due to the small magnitude of *P*. Recently, a theoretical study suggests that the single-phase hexagonal $LuFeO_3$ (*h*-$LuFeO_3$) system may have the potential to realize this ideal effect with the direct domain coupling between its ferroelectricity (FE) and weak ferromagnetism (WFM) [10]. Thus, it is of great importance to study experimentally the possibility of a direct domain coupling effect in $h$-LuFeO$_3$, which is energy efficient and highly desirable for future applications. Meanwhile, the recent ME coupling study on $h$-LuFeO$_3$/LuFe$_2$O$_4$ superlattices [11] demonstrates the capability of electrical field control of magnetism near the room temperature, which indicates $h$-LuFeO$_3$ and its related compounds are promising for future applications. However, with little knowledge of the intrinsic ME coupling of $h$-LuFeO$_3$ itself, a full understanding of the ME coupling in superlattices and other related materials is unrealistic.

For the possible giant ME coupling, $h$-LuFeO$_3$ systems in the P6$_3$cm polar structure (**Figure 1a**) has attracted a significant research attentions, since this metastable hexagonal phase of its orthorhombic bulk form was reported to be stabilized in thin film form by epitaxial strain [12,13] or in bulk by Sc or Mn doping [14–16]. Samples in both forms are believed to show similar physical properties despite different approaches used for the synthesis (see Supplementary Information, note 1 for details). They exhibit robust ferroelectricity well above room temperature, similar to the well-studied multiferroic hexagonal RMnO$_3$ ($h$-RMnO$_3$, R=rare earth) [17,18]. Their improper ferroelectricity due to the structural instability is associated with the structural trimerization. The trimerization drives the FE transition from the non-polar P6$_3$/mmc to the polar P6$_3$cm structure and results in a two-up/one-down (or two-down/one-up) displacement of R-site ions, which, in turn, leads to ferroelectric polarization along the $c$ axis. Note that alternating FE domains are clamped to the six structural vortex domains around one topological defect, forming an intriguing cloverleaf pattern in $h$-RMnO$_3$ (ref. 19). Although magnetic domains in $h$-RMnO$_3$ can be coupled with FE domains [20,21], there exists no bulk spontaneous $\boldsymbol{M}$ in $h$-RMnO$_3$.

Compared with $h$-RMnO$_3$, the strong magnetic interaction between Fe$^{3+}$ spins in $h$-LuFeO$_3$ induces a high antiferromagnetic ordering temperature ($T_N$= 440 K in ref. 12, 170 K in ref. 13). Even though the exact $T_N$ is still under debate, the presence of net ferromagnetic moment along the $c$ axis is evident below 160-170 K. The magnetic state below 160-170 K is confirmed in both theory and experiment to be the A$_2$-dominant spin configuration (Figure 1b) [10,15], which results in a significant canted-antiferromagnetic (or WFM) moment along the $c$ axis. Both large spontaneous $\boldsymbol{P}$ and $\boldsymbol{M}$ do coexist below 160-170 K. More importantly, a theoretical study predicts the presence of a bulk linear ME coupling or even the direct coupling between polarization and WFM domains in $h$-LuFeO$_3$ (ref. 10). Therefore, $h$-LuFeO$_3$ is a unique candidate where a direct mutual control of $\boldsymbol{P}$ and $\boldsymbol{M}$ domains can be explored.

**Results**

Despite all of these extensive investigations, the intrinsic FE domain structure in $h$-LuFeO$_3$ is still unexplored, and the direct experimental study of the ME coupling in $h$-LuFeO$_3$ is absent. One of the main reasons is the instability of the $h$-LuFeO$_3$ phase at ambient synthesis conditions, which makes it challenging to synthesize bulk single crystals. Here, we report a successful growth of high-quality highly-cleavable $h$-Lu$_{1-x}$Sc$_x$FeO$_3$ (nominal x=0.4) single crystals for the first time, using a floating zone technique and subsequent extensive annealing with different cooling rates (see Methods and Supplementary Information, note 2). The X-ray spectrum obtained using a PANalytical diffractometer (Figure 1c) is consistent with the pure $h$-Lu$_{0.6}$Sc$_{0.4}$FeO$_3$ phase [14,16], and does not exhibit any measurable peak broadening or the presence of any second phases, indicating the high quality of crystals. Three 1 °C/h-cooled $h$-Lu$_{0.6}$Sc$_{0.4}$FeO$_3$ specimens (LSFO1-3) were prepared by mechanical cleaving to expose hexagonal $a$-$b$ surfaces, and used for most of our scanning experiments. A large and flat surface of LSFO1 after cleaving is shown in Figure 1d, while an image of the whole as-grown crystal is shown in the inset. From room-temperature piezoresponse force microscopy (PFM) studies, we have visualized a cloverleaf pattern of vortex FE domains in $h$-Lu$_{0.6}$Sc$_{0.4}$FeO$_3$ for the first time, which is absent in epitaxial $h$-LuFeO$_3$ thin films. Furthermore, low-temperature magnetic force microscopy (MFM) studies on the same surfaces suggest that WFM domains are distinct from FE domains in terms of the size and shape, and there exists no mutual locking between FE and WFM domain walls. These observations undoubtedly indicate a complete decoupling between FE and WFM.

### A. Ferroelectricity

Figure 2a, 2b, and 2c are the PFM images of cleaved $h$-Lu$_{0.6}$Sc$_{0.4}$FeO$_3$ surfaces with different cooling rates in the 1200 °C-1400 °C range, in which FE Curie temperature ($T_C$) locates (see Supplementary Information, note 2). The fast-cooled sample (100 °C/h) shows small disordered FE domains on a ~100 nm scale (Figure 2a). Additional dark field transmission electron microscopy (DF-TEM) studies confirm that a 100 °C/h-cooled $h$-Lu$_{0.5}$Sc$_{0.5}$FeO$_3$ crystal also exhibits irregular-shape and disordered FE domains on a similar scale (Supplementary Information, Figure S1), possibly due to a large amount of chemical/structural disorders (e.g. partial edge dislocations [22] and large line defects [23] ). A cloverleaf pattern of vortex FE domains, similar to that in $h$-RMnO$_3$ (ref. 19), is now visible in the 10 °C/h-cooled specimen (Figure 2b). A well-organized micron-size cloverleaf pattern of vortex FE domains (Figure 2c and 2d) is evident in the 1 °C/h-cooled LSFO1 (area A). We found that vortex FE domains distribute uniformly across the whole cleaved surface. For example, Figure 2f shows the PFM image of another area (area B, far away from area A) of LSFO1, resembling those of area A. These results demonstrate that the intrinsic FE domain structure in $h$-Lu$_{0.6}$Sc$_{0.4}$FeO$_3$ is a topological vortex configuration, which is sensitive to chemical/structural disorders. Their relevant length scale can be systematically tuned by the cooling rate across the FE Curie temperature, which is in the range of 1400 °C-1200 °C. Although most of our samples have FE domains with +P and −P domains in 50/50 ratio,

known as type-I domains, some cleaved samples closed to the surface of crystals do show type-II narrow domains (Figure S2) due to the self-poling effect during the annealing process, which is also commonly observed in $h$-RMnO$_3$ compounds [24].

We plotted the density of topological defects (vortices and antivortices) in PFM images vs. the cooling rate (Figure 2e). The $h$-Lu$_{0.6}$Sc$_{0.4}$FeO$_3$ tends to show a much higher density of topological defects, compared with $h$-RMnO$_3$ (e.g. ErMnO$_3$ and TmMnO$_3$, data obtained from ref. 25), even though it has a higher $T_C$ than, e.g., ErMnO$_3$ ($T_C \approx 1129$ °C). Note that a higher-$T_C$ $h$-RMnO$_3$ tends to exhibit a lower density of topological defects. The estimated linear slope of ~0.88 is significantly larger than the slope (~0.59) predicted by the Kibble-Zurek mechanism (KZM) in $h$-RMnO$_3$ (ref. 25). This discrepancy is likely due to the high density of chemical/structural disorders in fast-cooled $h$-Lu$_{0.6}$Sc$_{0.4}$FeO$_3$, which can be annealed away through the slow cooling at the high-temperature range. This disorder is probably related to the metastable nature of the hexagonal phase of LuFeO$_3$, unlike the stable $h$-RMnO$_3$ phase. Meanwhile, the poor R-Square value (~0.84) of the linear fitting in $h$-Lu$_{0.6}$Sc$_{0.4}$FeO$_3$ (Figure 2e) also suggests that the density of topological defects is likely affected by extrinsic effects such as pinning by structural defects. Therefore, the intrinsic slope of $h$-Lu$_{0.6}$Sc$_{0.4}$FeO$_3$ needs to be further studied in the future, especially in the slowly-cooled range where the pinning effect by structural defects are minimized. We emphasize that we have observed the cloverleaf-pattern FE domains in $h$-LSFO, which has not been reported before, e.g., in epitaxial $h$-LuFeO$_3$ thin films [12]. Note that $h$-LuFeO$_3$ films are typically synthesized at 700 °C-900 °C, which is well below the $T_C$. Thus, it is possible that films may have significant chemical/structural disorders, resulting in ultra-fine irregular FE domains or single FE domain due to surface boundary conditions.

We obtained good polarization loops (Figure 2g) of another 1 °C/h-cooled specimen (LSFO2) using a Ferroelectric Material Test System (FMTS, Radiant), demonstrating the presence of switchable ferroelectric polarization. Complimentary polarization loops as a function of the frequency are also provided in the Supplementary Information Figure S3a. Their consistency and similarities at different frequencies suggest that the measured loops reflect their intrinsic polarizations without artifacts. The permittivity and the loss tangent as a function of frequency and temperature are shown in Figure S3b and S3c respectively. The low loss at the room temperature and its continuous drop at low temperatures indicate $h$-$Lu_{0.6}Sc_{0.4}FeO_3$ is a good insulator.

## B. Magnetism

The magnetic susceptibility of LSFO2 in magnetic fields along and perpendicular to the $c$ axis (Figure 3a), measured by a Magnetic Property Measurement System (Quantum Design), demonstrate the existence of the canted antiferromagnetic (or WFM) moment along the $c$ axis below 163 K, which is consistent with the presence of the $A_2$ spin order. A smooth reduction of magnetic susceptibility below 75 K upon cooling may indicate the presence of a spin reorientation transition away from the $A_2$ spin order, which has been reported in $Lu_{0.5}Sc_{0.5}FeO_3$ (ref. 15). This behavior is also evident in the M-H curves at 5 K and 130 K (Figure 3b). Thus, WFM domain structures were examined by a low-temperature MFM (Attocube) at the liquid nitrogen temperature (~78 K), where the $A_2$ phase (thus, WFM moment) is dominant. Compared with the micron-size vortex FE domains, large loop-like WFM domains (Figure 3c) without any hint of cloverleaf cores are clearly observed in LSFO1 (area C, Figure 3d). To further confirm the magnetic origin of these MFM contrasts, the specimen was warmed up to 200 K (above the $A_2$-phase transition temperature of 163 K) and cooled back to 78 K. After this new thermal cycle, the WFM domain structure of the same area has completely changed (Figure 3e and 3f). For a clear view of WFM domain structure, a large-range mosaic MFM image around the area C is shown in Figure 3g. The vortex FE domains at the lower-left corner of Figure 3g, reproduced from Figure 2d, are distinct from the large WFM domains. The size of these WFM domains is roughly ~100 μm, while FE domains are on the order of ~1 μm. Furthermore, the WFM domain wall thickness is as large as 2-4 μm (Figure S4), which is also distinct from the almost-atomically-sharp FE domain walls.

Moreover, our in-situ PFM images at the room temperature (Figure S5a and S5b) and 78 K (Figure S5c and S5d) show an identical domain pattern at both temperatures. Considering that WFM domains change in different thermal cycles (Figure 3c and 3e) while FE domains should remain unchanged, we could confidently draw the conclusion that they are decoupled.

### C. ME coupling

There are two and only two possible types of coupling among ferroelectric polarization, structural antiphases, and $A_2$-type antiferromagnetic spins that have been theoretically proposed [10]. They may result in different relationships among FE ($P= \pm P$), WFM ($M_c= \pm M_c$, magnetization along the $c$ axis), and ME ($\alpha_c= \pm\alpha_c$, ME coefficient along the $c$ axis) domains, where $P=\alpha_c \bullet M_c$. First, note that within unit-cell translation, structural distortions such as oxygen distortions rotate $|\pi/3|$ across one FE domain wall, which is a structural antiphase boundary at the same time, so that the six domain structure in Figure 4 is a topological vortex. Now, in the first case (Figure 4a), in-plane spins rotate $|\triangle\Phi| = 2\pi/3$ across one FE domain wall. In this case, FE domains are directly coupled with WFM domains but decoupled from ME domains. So WFM domains and FE domains must exhibit an identical pattern. In the other case (Figure 4b), in-plane spins rotate $|\triangle\Phi| = \pi/3$ across one FE domain wall, which is in the same manner as structural distortions. So FE domains will be decoupled with WFM domains and coupled with ME domains. Our experimental findings clearly indicate the decoupling between FE and WFM domains (Figure 4b). It can be summarized in the cartoon of Figure 4c, obtained by combining Figure 2d and 3e. Note that similar PFM and MFM results have been reproduced in another cleaved crystal (LSFO3) of the same batch (Figure S6). Moreover, the similar magnetic susceptibility and PE loop can also be reproduced in different crystal pieces (Figure S7) that are cleaved from the same batch, which is an indication of uniform ferroelectric and magnetic properties within the batch. Figure 4d illustrates the zoom-in view of the dotted area in Figure 4c, and the 3D spin configurations of trimerized $Fe^{3+}$ are drawn for each domain. Another evidence of this domain decoupling between FE and WFM is the absence of magnetoelectric current (or $P$ switching) when $M_c$ is switched by magnetic fields (see Figure S8 and Supplementary Information note 3 for details).

**Discussion**

Our results of decoupled FE and WFM domains can be understood in terms of magnetic domain wall energy. It is expected that the domain wall energy with the in-plane spin rotation of large $|2\pi/3|$ likely costs larger than that with the small $|\pi/3|$ rotation. This can also be understood in terms of topology. Structural distortions such as oxygen distortions all the way around one vortex (antivortex) core rotate by $+(-)2\pi$, so the relevant structural topological charge (or the winding number, $n_s$) of one vortex (antivortex) is $+(-)1$. Similarly, magnetic topological charge (or the winding number, $n_m$) for the in-plane spins of a vortex (antivortex) in the domain-decoupled case (Figure 4b) is $+(-)1$. On the other hand, $n_m$ is $+(-)2$ in the domain-coupled case (Figure 4a), which is different from $n_s$ and likely not favored topologically. As a consequence of this coupling between structural distortions and in-plane antiferromagnetic spins, the out-of-plane spins (WFM) are decoupled with FE domains as observed experimentally. Though a further experimental proof of this coupling is needed in the future, the statement above is still logically true as the coupling between structural distortions and in-plane antiferromagnetic spins is the only possibility based on theoretical predictions (ref. 10). This is also consistent with the situation in $YMnO_3$ where FE domains are found to be coupled to its in-plane antiferromagnetic domains [21]. We want to emphasize that the spin-canting angle in $h$-$Lu_{0.6}Sc_{0.4}FeO_3$, estimated from the observed remnant magnetization (~ 0.01 $\mu_B$/Fe), is tiny (< 1 °). Therefore, antiferromagnetic in-plane spins are still dominant. So it is natural to have $n_m=n_s$, which leads to the decoupling between FE and WFM domains, since WFM is only a small out-of-plane spin component of the dominant in-plane spins.

In summary, we have found that the intrinsic FE domain structure of $h$-LSFO is of fine topological vortex configuration. In contrast, weak ferromagnetic domains are of of a loop shape with a much larger size. The distinct sizes and shapes demonstrate the decoupling between ferroelectricity and out-of-plane weak ferromagnetism, even though in-plane antiferromagnetic spins likely correlate with structural modulations around topological vortices according to theoretical predictions. Our results indicate that the magnetic topological charge tends to be identical with the structural topological charge. This observation is consistent with the absence of any FE and WFM domain-coupled ME effects in $h$-LSFO. Our results provide new insights into induction of direct coupling between magnetism and ferroelectricity mediated through structural distortions, which will be useful for the future applications of multiferroics.

**Methods**

**Crystal growth**

High-quality $h$-$Lu_{0.6}Sc_{0.4}FeO_3$ single crystals were grown using a floating zone method under 0.8 MPa $O_2$ atmosphere. The feed rods were prepared using the standard solid-state reaction (ref. 14). The as-grown crystals were annealed at 1400 ℃ in air for 24 hours to enhance crystallinity, and then cooled down to 1200 °C with different cooling rates (100 °C/h, 10 °C/h and 1 °C/h). Afterward, they were cooled to room temperature at the same cooling rate of 100 °C/h. Finally, they were further annealed under 20 MPa $O_2$ pressure at 1000 °C in a high-pressure oxygen furnace to remove any oxygen vacancies and thus enhance resistance. In similar $h$-RMnO$_3$ compounds, it is well known that the topological vortex density can be symmetrically tuned by different cooling rates near their FE Curie temperatures (ref. 23). As different densities of vortex FE domains are observed in samples with different rates for the 1200 °C-1400 °C cooling in this work, the FE Curie temperature of $h$-Lu$_{0.6}$Sc$_{0.4}$FeO$_3$ is in the range of 1200 °C-1400 °C.

### PFM and MFM measurements

For all the PFM measurements, AC 5V at 68 kHz was applied to a tip (a commercial conductive AFM tip) while sample backside is grounded. Before the MFM measurement, a 20 nm gold film was sputtered onto the same cleaved surface of LSFO1 after PFM experiments to eliminate any electrostatic signal from FE domains during the MFM scanning (ref. 20). All the MFM images are obtained using a dual pass mode with a lift height of 50 nm.

### PE measurements

PE loops were measured on a polished thin sample by the ''PUND'' method provide in the Ferroelectric Material Test System (RADIANT TECHNOLOGIES INC.). The remanent-only polarization hysteresis loop is finally derived. (Also see the reference in Supplementary information note 2)

### Data availability

All relevant data are available from the authors upon request.

**Acknowledgements**

We would like to thank M. Mostovoy for insightful discussions. This work was supported by the Gordon and Betty Moore Foundation's EPiQS Initiative through Grant GBMF4413 to the Rutgers Center for Emergent Materials.

**Competing interests**

The authors declare no competing interests.

**Contributions**

B. G.,K.D., X. X. and R.H. prepared the samples; F. H. performed the TEM investigation; Y. W. and J. K. measured P-E loops and magnetic properties; K. D. did PFM and MFM measurements; S. C. guided the project; K.D. and S. C. analysed the data and wrote the paper.

**Supplementary Information**

Supplementary information is available online or from the author.

**Figure legends**

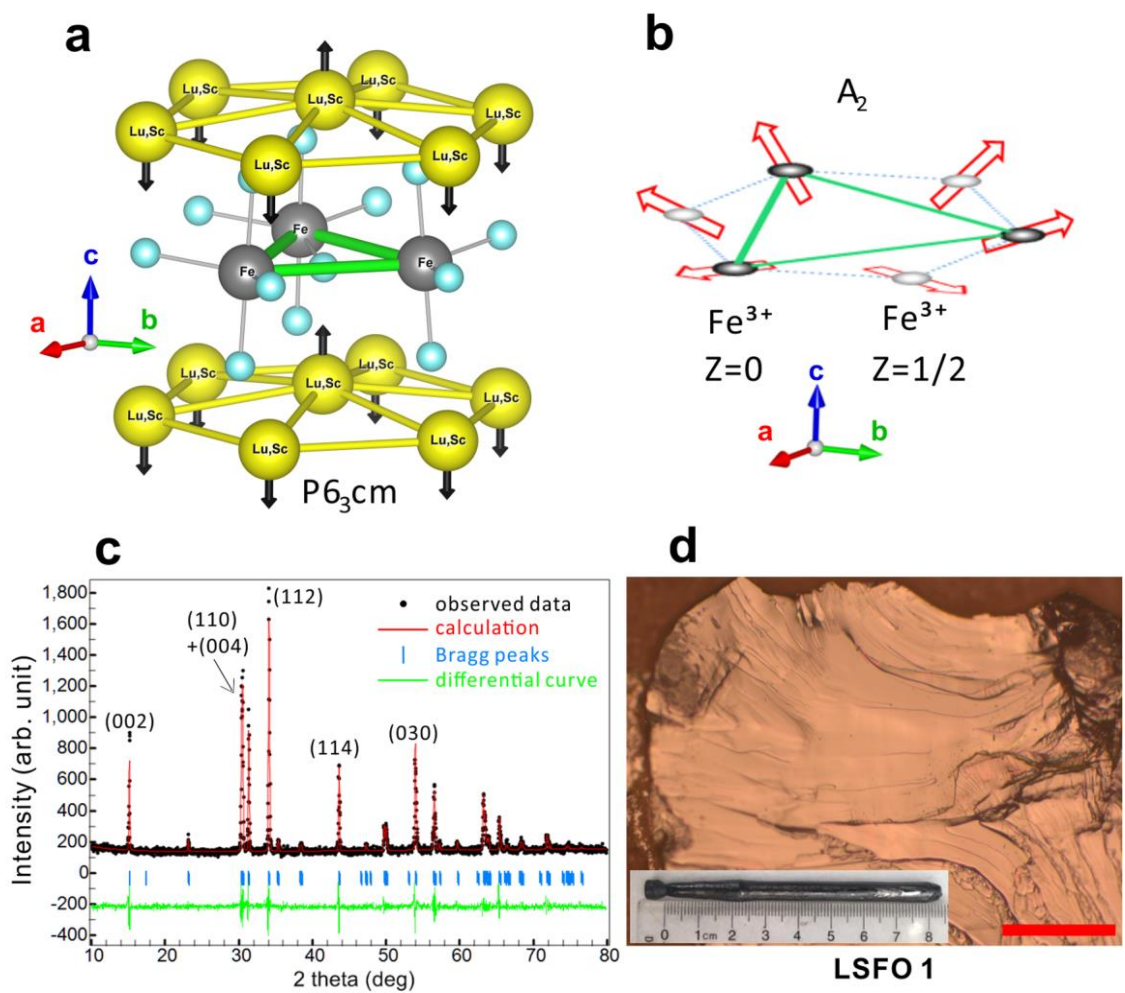

**Figure 1: The structure of *h*-(Lu, Sc)FeO₃.** **(a)** Crystallographic structure of *h*-(Lu, Sc)FeO₃ in P6₃cm space group**.** Yellow balls indicate Lu or Sc atoms, while grey and blue balls stand for Fe and O atoms respectively. The black arrows show the displacement directions of Lu or Sc atoms with a two-down/one-up case. Trimerized Fe atoms are illustrated with green bonds. **(b)** 3D magnetic structure of *h*-(Lu, Sc)FeO₃ in A₂ phase, which enables the spin canting along the *c* axis. Ordered Fe spins in two adjacent layers are shown with red open arrows, and the Fe trimerization of Z=0 layer is labeled with green bonds. **(c)** X-ray diffraction of the Lu₀.₆Sc₀.₄FeO₃ single crystal ground powders which shows the pure hexagonal phase in P6₃cm space group. The refinement was done by Reitica with the simulated lattice constant a=5.879 Å and c= 11.695 Å, the standard deviation Rp=6.851, Rwp=10.62 and χ2=0.514 **(d)**

Optical microscope image of a cleaved Lu$_{0.6}$Sc$_{0.4}$FeO$_3$ surface (LSFO1). Scale bar is 200 μm. The image of the whole crystal before cleaving is shown in the inset.

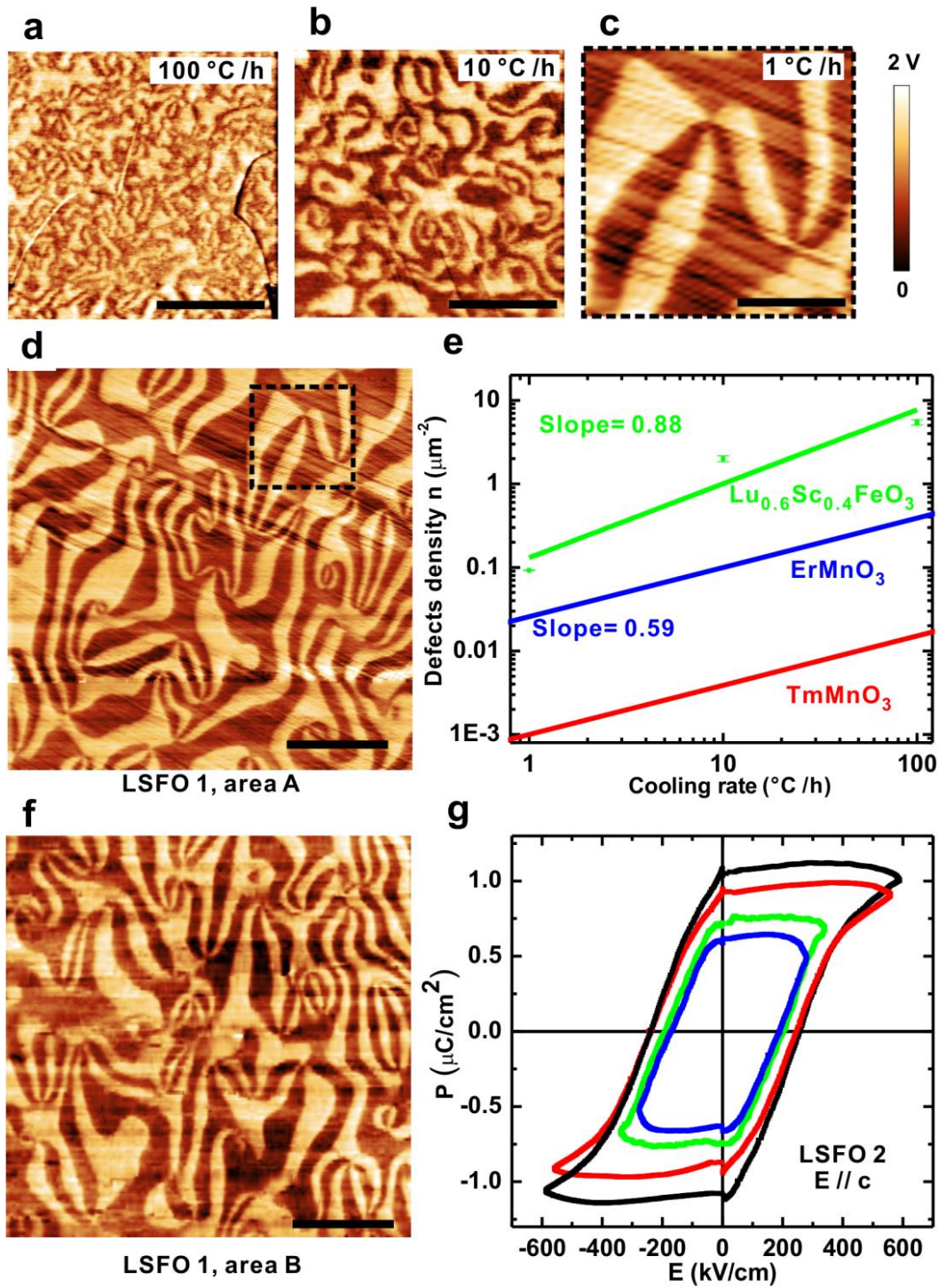

**Figure 2: The ferroelectricity and vortex ferroelectric domains of $h$-Lu$_{0.6}$Sc$_{0.4}$FeO$_3$.** Room-temperature PFM images of cleaved surfaces of crystals after **(a)** 100 ℃/h, **(b)** 10 ℃/h and **(c)** 1 ℃/h cooling in the 1400 °C-1200 °C temperature range. **(d)** The large-range PFM image of the area A of LSFO1 (the black dashed square corresponds to (c)). **(e)** The density of defects as a function of cooling rates. Error bars of the defect density for 1 ℃/h, 10 ℃/h, and 100 ℃/h cooling are ± 0.005 μm$^{-2}$, ± 0.16 μm$^{-2}$, and ±0.4 μm$^{-2}$ respectively. Data of ErMnO$_3$ and TmMnO$_3$ is obtained from ref. 25. **(f)** PFM image of another area in LSFO1 (area B). **(g)** Room-temperature P-E loops of LSMO2 at 2702 Hz, which prove its robust and switchable ferroelectricity. Scale bars are 2 μm in (a), (b) and (c), 5 μm in (d) and (f).

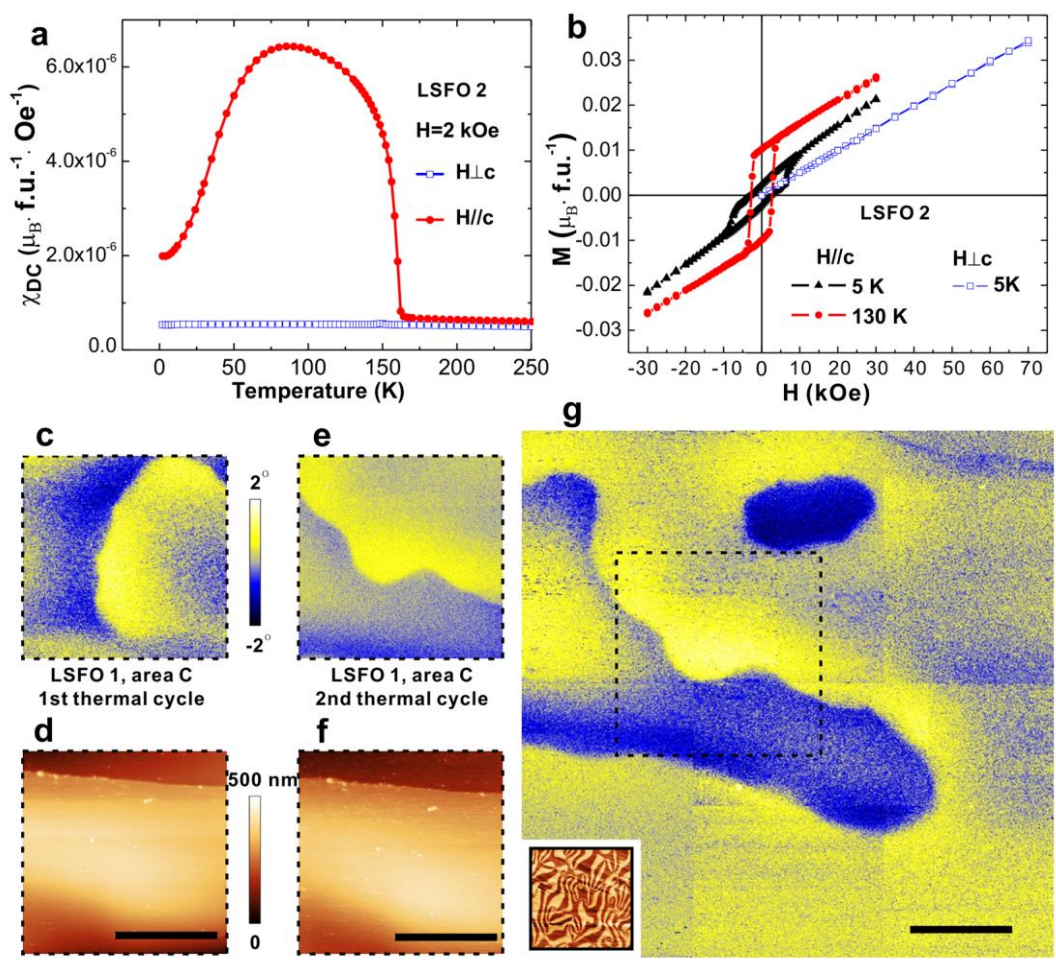

**Figure 3: Magnetic properties and magnetic domains of *h*-Lu$_{0.6}$Sc$_{0.4}$FeO$_3$.** (a) magnetic susceptibility of LSFO2 in 2 kOe (H⊥c; open squares, and H//c; red circles) as a function of temperature. (b) Out-of-plane magnetization (H//c) of LSFO2 as a function of magnetic field at 130 K (red circles) and 5 K (black triangles), and linear in-plane magnetization (H⊥c) at 5 K (blue open squares). (c) MFM image and (d) corresponding topography of LSFO1 (area C) at 78 K. (e) MFM image and (f) its topography of the same area C at 78 K after a new thermal cycle. (g) Large-range MFM image collage around the area C, which is shown with the black dashed square. PFM image of LSFO1 (area A) is shown in the black square at the left-lower corner with the same scale for comparison. Large loop-shape WFM domains are clearly visible in ~100 μm scale. Scale bars are 20 μm.

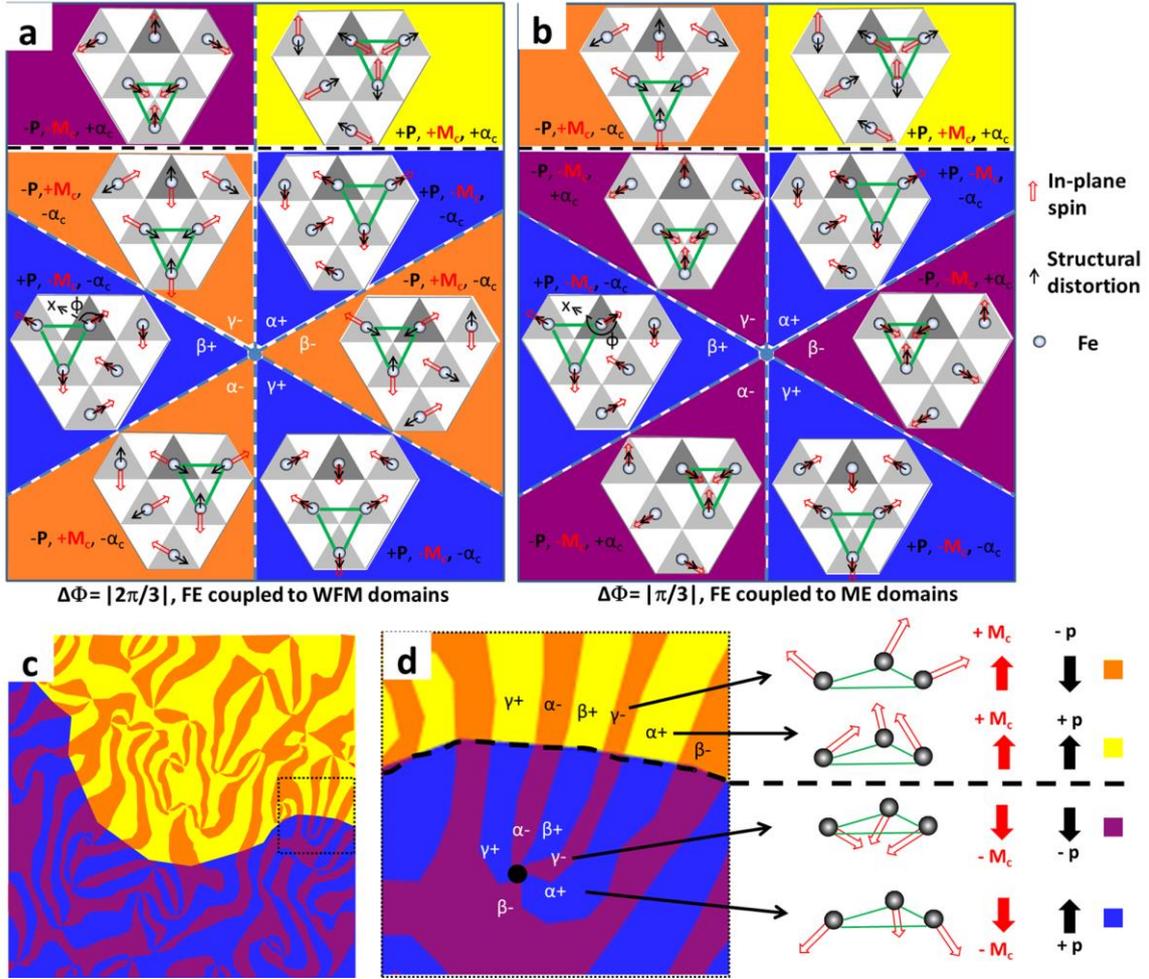

**Figure 4: Magnetoelectric coupling of *h*-Lu$_{0.6}$Sc$_{0.4}$FeO$_3$.** Possible in-plane magnetic structures of coupled **(a)** and decoupled **(b)** cases between FE and WFM domains. Blue (black) dashed lines show the FE (WFM) domain walls. Dark circles are Fe$^{3+}$ ions, and the red open arrows illustrate their in-plane spins. Black arrows stand for the structural distortions of top-apical oxygen ions. Trimerized Fe$^{3+}$ ions are labeled with green bonds. The blue (+***P***, -***M**_c*), yellow (+***P***, +***M**_c*), orange (-***P***, +***M**_c*) and purple (-***P***, -***M**_c*) background colors represent the 4 possible FE and WFM domain combinations. ***α**_c* is the ME coefficient along the *c* axis. The α+ domain is chosen to be the same for clear comparison. **S**pins in dark grey triangle backgrounds are supposedly identical by unit-cell translations, rotate |Δ Φ| = 2π/3 for **(a)** and |Δ Φ| = π/3 for **(b)** across an FE domain wall. **(c)** Cartoon of real FE and WFM domain distributions drawn by combining Fig. 2d and 3e in the same scale. **(d)** Zoom-in cartoon of the black dotted area in **(c)**, and the 3D cartoon of spin configurations for four different domains. Spins are shown by the red open arrows, and other symbols remain the same as in **(a)** and **(b)**, except that red (black) solid arrows are now the $c$-direction canted magnetic moments (electric polarizations). The domain decoupled case of **(b)** corresponds to the experimental result of **(c)** or **(d)**, which demonstrates decoupling between FE and WFM domains.

# Supplementary Information

**Title**

## Vortex Ferroelectric Domains, Large-loop Weak Ferromagnetic Domains, and Their Decoupling in Hexagonal (Lu, Sc)FeO$_3$

*Kai Du, Bin Gao, Yazhong Wang, Xianghan Xu, Jaewook Kim, Rongwei Hu, Feiting Huang and Sang-Wook Cheong\**

Rutgers Center for Emergent Materials and Department of Physics and Astronomy, Rutgers University, Piscataway, New Jersey 08854, USA

\* To whom the correspondence should be addressed. (E-mail: sangc@physics.rutgers.edu)

### Note 1: Difference between Sc-doped *h*-LuFeO$_3$ and pure *h*-LuFeO$_3$

Hexagonal rare-earth ferrites (*h*-RFeO$_3$) are metastable and do not exist under normal synthesis conditions, unlike the stable hexagonal rare-earth manganites. In order to stabilize the hexagonal phase experimentally, strain from small-ion doping into A-site for bulk crystals (e.g. *h*-Lu$_{0.6}$Sc$_{0.4}$FeO$_3$) or from the hexagonal substrate for thin films (e.g. *h*-LuFeO$_3$/YSZ) are required. Generally, these two approaches apply strain to its original stable lattice and should give the similar effect in stabilizing the metastable hexagonal phase.

Besides, strong theoretical and experimental evidence has illustrated that Sc-doped *h*-LuFeO$_3$ and pure *h*-LuFeO$_3$ do share the similar ferroelectric and magnetic properties. Theoretically, DFT calculations have been conducted to the Sc-doped *h*-LuFeO$_3$ in ref. 16 compared to the pure *h*-LuFeO$_3$. It concludes that the Sc doping will not affect the ferroelectricity and magnetic properties, other than just helping to stabilize the hexagonal phase. A secondary effect of the small Sc doping is: Fe-Fe distance becomes closer, and thus their magnetic interaction becomes stronger, which is consistent with high magnetic transition temperatures of the Sc-doped *h*-LuFeO$_3$.

Experimentally, it is also true that our $h$-$Lu_{0.6}Sc_{0.4}FeO_3$ has a slightly higher weak ferromagnetic transition temperature at 160 K-170 K, comparing to pure $h$-$LuFeO_3$ thin films (140 K-150 K). Except for that small difference, bulk Sc-doped $h$-$LuFeO_3$ shows similar magnetic properties to pure $h$-$LuFeO_3$ thin films, based on the magnetic susceptibility and neutron diffraction study in ref. 13 and ref. 15.

Therefore, $h$-$Lu_{0.6}Sc_{0.4}FeO_3$ in this work should have the similar properties to theoretically studied pure $h$-$LuFeO_3$, based on both first principle calculations and the experimental data. However, as mentioned later in the main text, our bulk $h$-$Lu_{0.6}Sc_{0.4}FeO_3$ single crystals have much better quality so that it could enable us to further study their intrinsic FE domains and ME coupling, which is difficult to realize in pure $h$-$LuFeO_3$ thin films. Thus, bulk Sc-doped $h$-$LuFeO_3$ can be a better platform to explore the intrinsic properties and novel fundamental physics in the $h$-$LuFeO_3$ system.

**Note 2: Crystal growth and preparations for measurements**

High-quality $h$-$Lu_{0.6}Sc_{0.4}FeO_3$ single crystals were grown using a floating zone method under 0.8 MPa $O_2$ atmosphere. The feed rods were prepared using the standard solid-state reaction (ref. 14). The as-grown crystals were annealed at 1400 ℃ in air for 24 hours to enhance crystallinity, and then cooled down to 1200 ℃ with different cooling rates (100 ℃/h, 10 ℃/h and 1 ℃/h). Afterward, they were cooled to room temperature at the same cooling rate of 100 ℃/h. Finally, they were further annealed under 20 MPa $O_2$ pressure at 1000 ℃ in a high-pressure oxygen furnace to remove any oxygen vacancies and thus enhance resistance. In similar $h$-$RMnO_3$ compounds, it is well known that the topological vortex density can be symmetrically tuned by different cooling rates near their FE Curie temperatures (ref. 23). As different densities of vortex FE domains are observed in samples with different rates for the 1200 ℃-1400 ℃ cooling in this work, the FE Curie temperature of $h$-$Lu_{0.6}Sc_{0.4}FeO_3$ is in the range of 1200 ℃-1400 ℃.

For all the PFM measurements, AC 5V at 68 kHz was applied to a tip (a commercial conductive AFM tip) while sample backside is grounded. Before the MFM measurement, a 20 nm gold film was sputtered onto the same cleaved surface of LSFO1 after PFM experiments to eliminate any electrostatic signal from FE domains during the MFM scanning (ref. 20). All the MFM images are obtained using a dual pass mode with a lift height of 50 nm.

PE loops were measured on a polished thin sample by the "PUND" method [1] provide in the Ferroelectric Material Test System (RADIANT TECHNOLOGIES INC.). The remanent-only polarization hysteresis loop is finally derived.

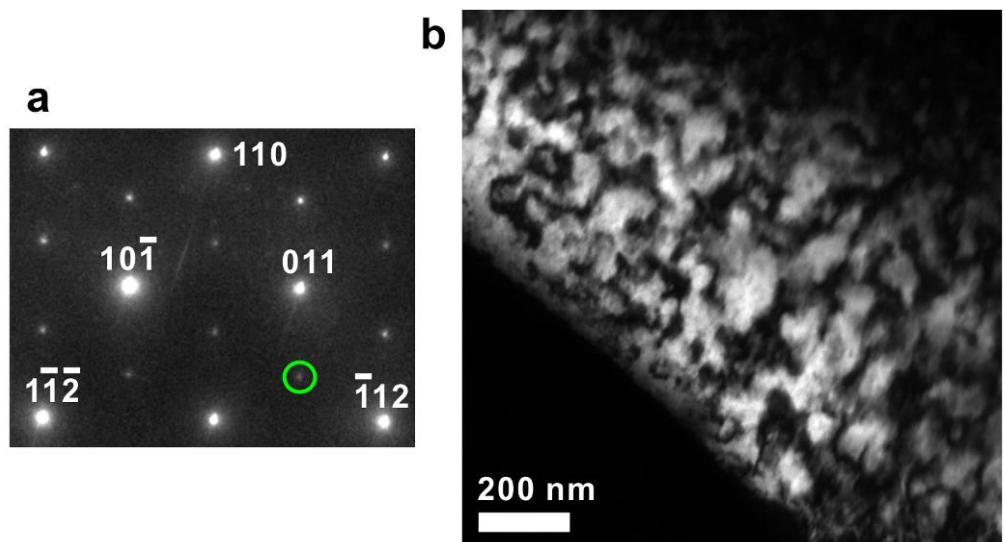

**Figure S1: Dark field transmission electron microscopy (DF-TEM) image of the fast-cooled (100 ℃/h) $Lu_{0.5}Sc_{0.5}FeO_3$.** (a) The diffraction pattern of the (1-11) surface. (b) FE domain image using the Friedel pair breaking of the ⅓(−123) superlattice spot (labeled with green circle). Small disordered and irregular FE domains with highly-curved domain walls are visualized.

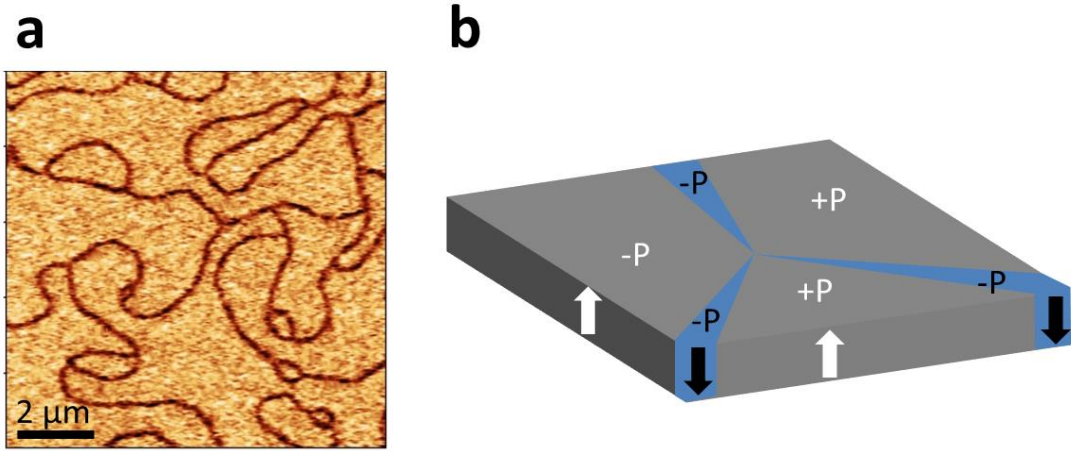

**Figure S2: Type-II narrow ferroelectric domains of Lu$_{0.6}$Sc$_{0.4}$FeO$_3$ due to the self-poling effect.** (a) PFM image of a cleaved surface closed to the surface of the crystal. (b) Cartoon of the type-II domains on a self-poled sample. Because of the oxygen gradient created near the sample surface during the annealing process, +P domains are poled to be dominant. –P domains contract and become narrow lines in shape.

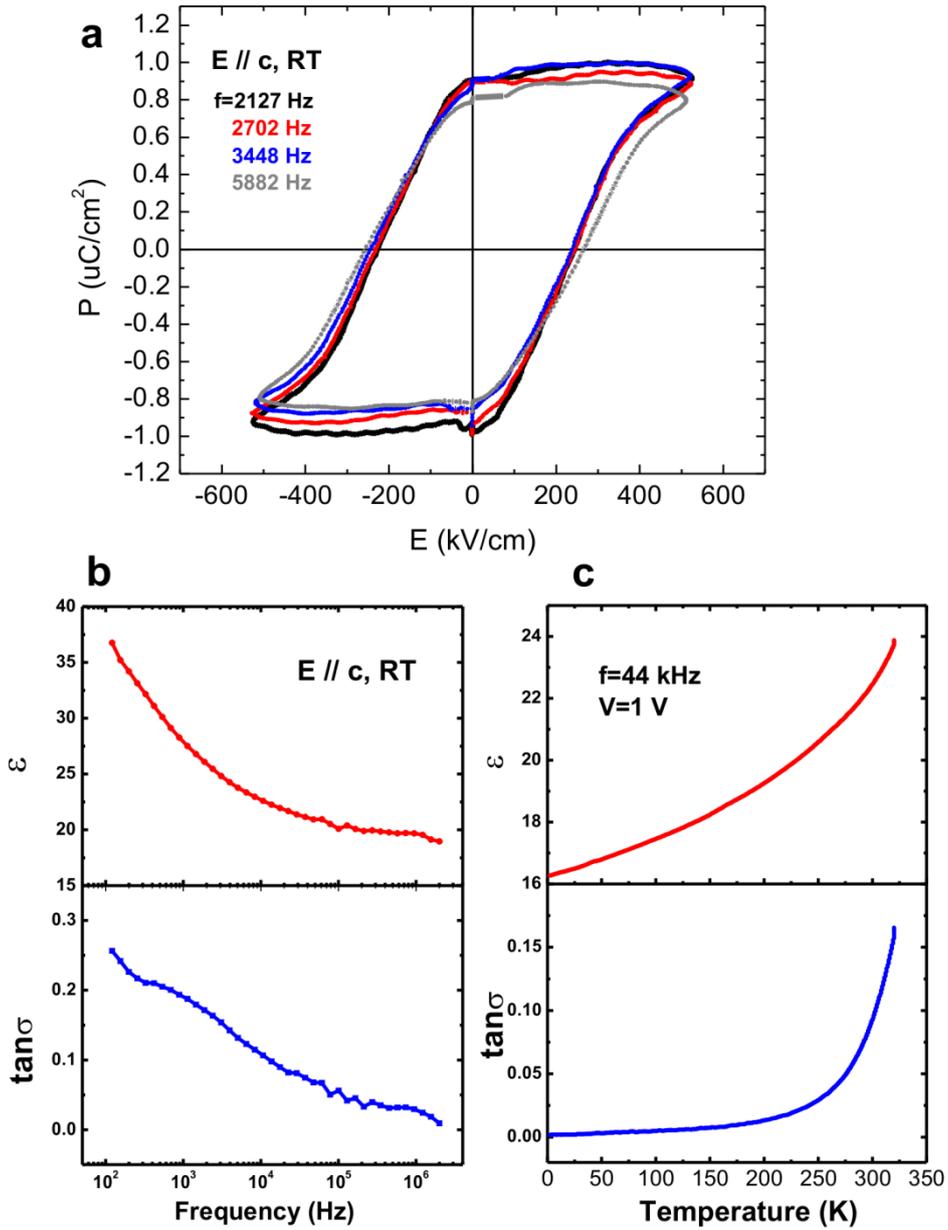

**Figure S3: Ferroelectric and dielectric properties of Lu$_{0.6}$Sc$_{0.4}$FeO$_3$.** (a) PE loops as a function of frequency. As there are no clear differences in PE loops with different frequencies, PE curves are from intrinsic polarizations other than leakage current contributions. (b) The real part of permittivity and the loss tangent as a function of frequency at the room temperature. (c) The real part of permittivity and the loss tangent as a function of temperature. The low loss at the room temperature and its continuous drop at low temperatures indicate the sample is a good insulator.

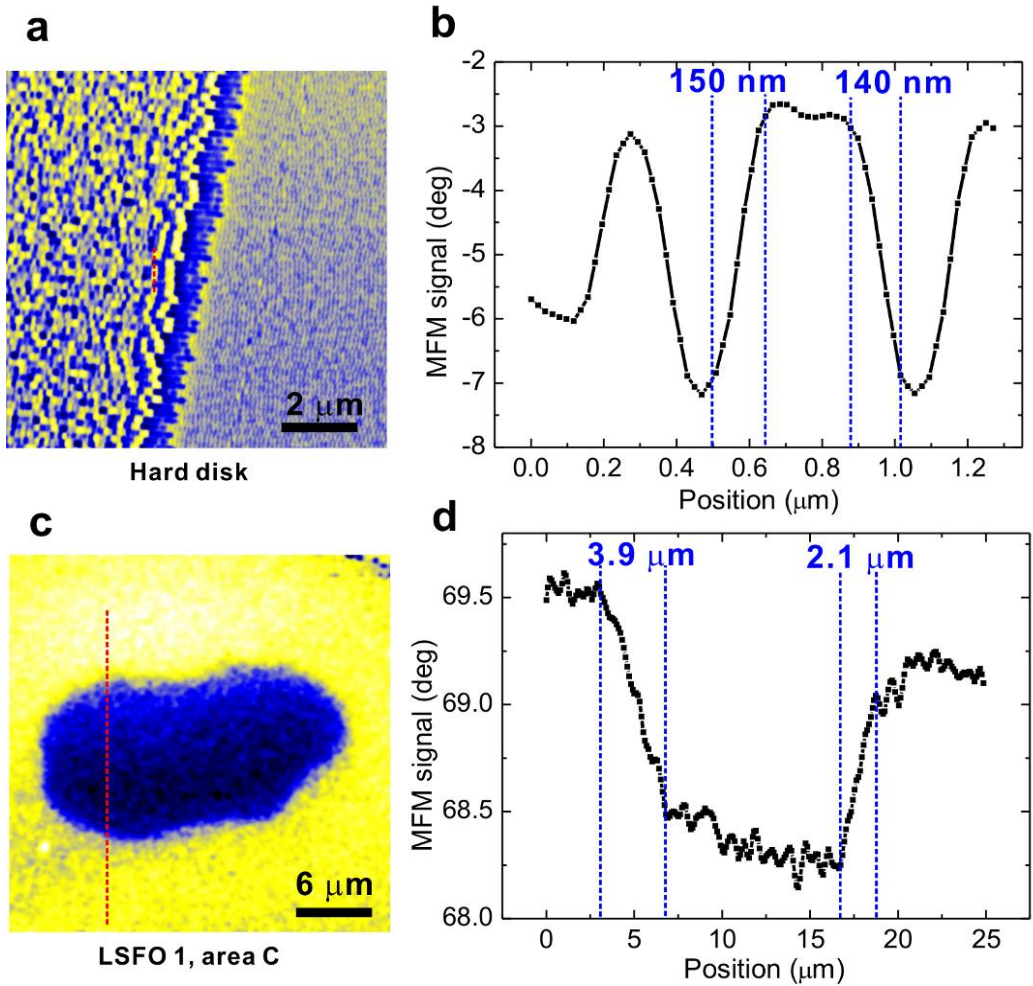

**Figure S4: WFM domain wall thickness. (a)** MFM image and **(b)** the line-profile of the red dashed line of a hard disk sample. 150 nm-sized domains are well resolved in the MFM image, which indicates that the spacial resolution of MFM is better than 150 nm. **(c)** MFM image and **(b)** the line-profile of the red dashed line of LSFO1 (area C). The same tip and experimental set-up were used for both experiments, and the domain wall thickness estimated from the MFM line-profile in **(d)** (the width of the pronounced steep lines) is larger than 2 μm.

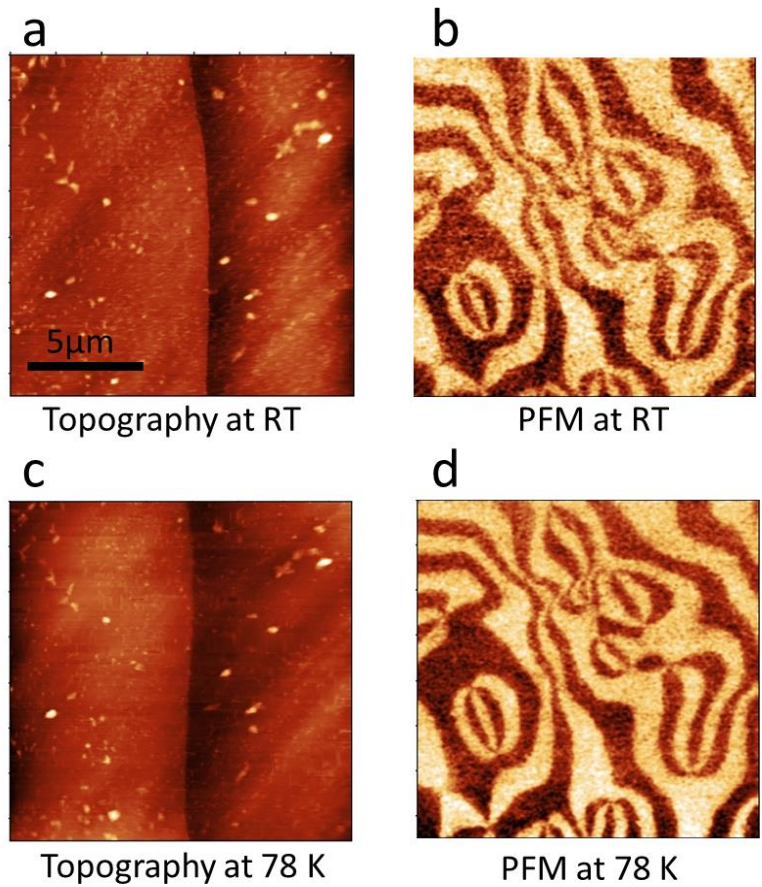

Topography at RT      PFM at RT

Topography at 78 K      PFM at 78 K

**Figure S5: FE domains at diferent temperatures** (**a**) Topography and (**b**) corresponding PFM image of LSFO3 (area C) at the room temperature. (**c**) Topography and (**d**) corresponding PFM image of LSFO3 (area C) at 78 K. Identical FE domains are observed. Tiny differences are due to the possible scanner distortions at different temperatures

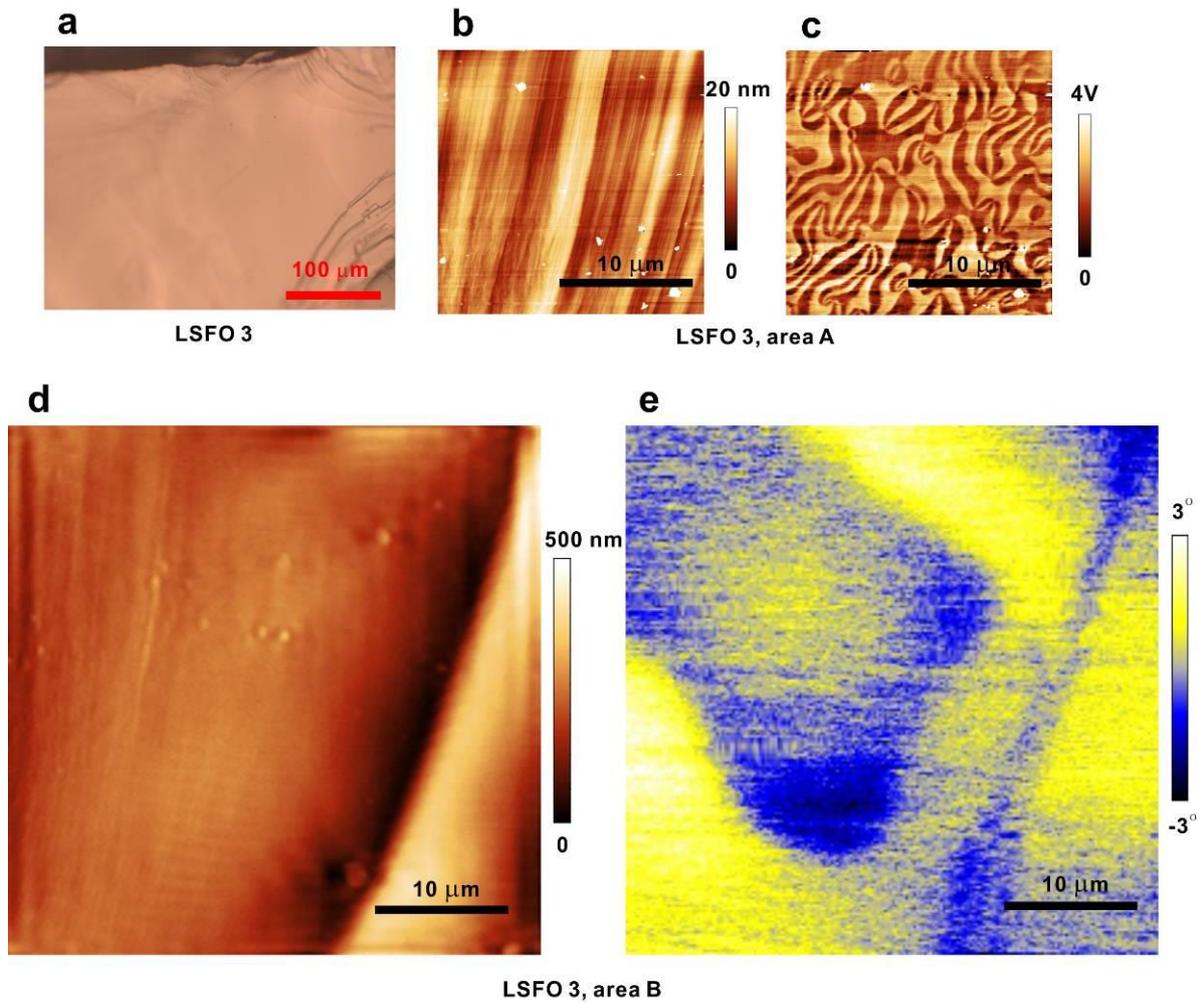

**Figure S6: Magnetoelectric domain decoupling in another piece of cleaved $h$-Lu$_{0.6}$Sc$_{0.4}$FeO$_3$ (LSFO3).** **(a)** Optical microscope image of the cleaved LSFO3. The red scale bar is 100 μm. **(b)** Topography and **(c)** corresponding PFM image of LSFO3 (area A) at the room temperature. **(d)** Topography and **(e)** corresponding MFM image of LSFO3 (area B) at 78 K. The black scale bar is 10 μm. FE and WFM domains form in very different length-scales.

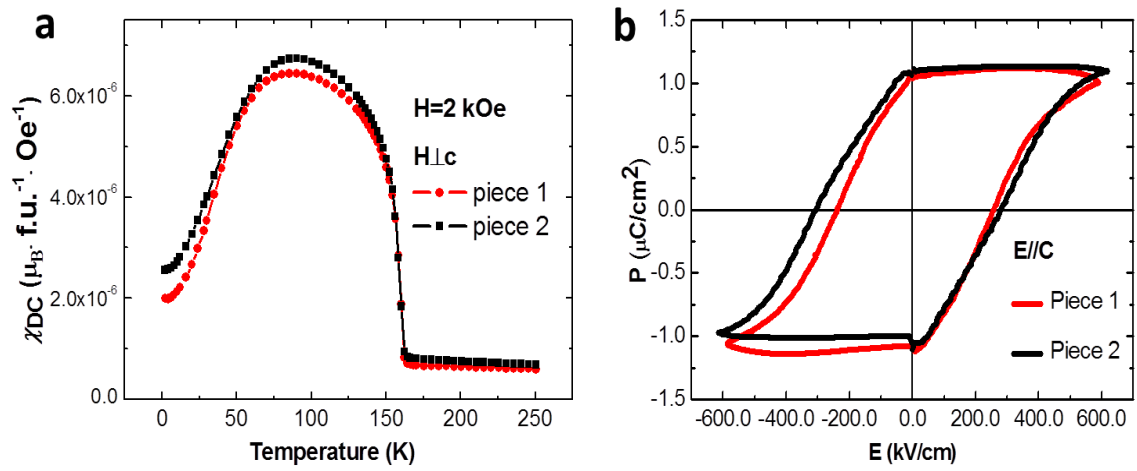

**Figure S7: Magnetic susceptibilities and PE loops of different crystal pieces cleaved from the same batch.** (**a**) Magnetic susceptibilities. (**b**) PE loops. Reproducible data indicates the ferroelectric and magnetic properties are uniform for samples from the same batch.

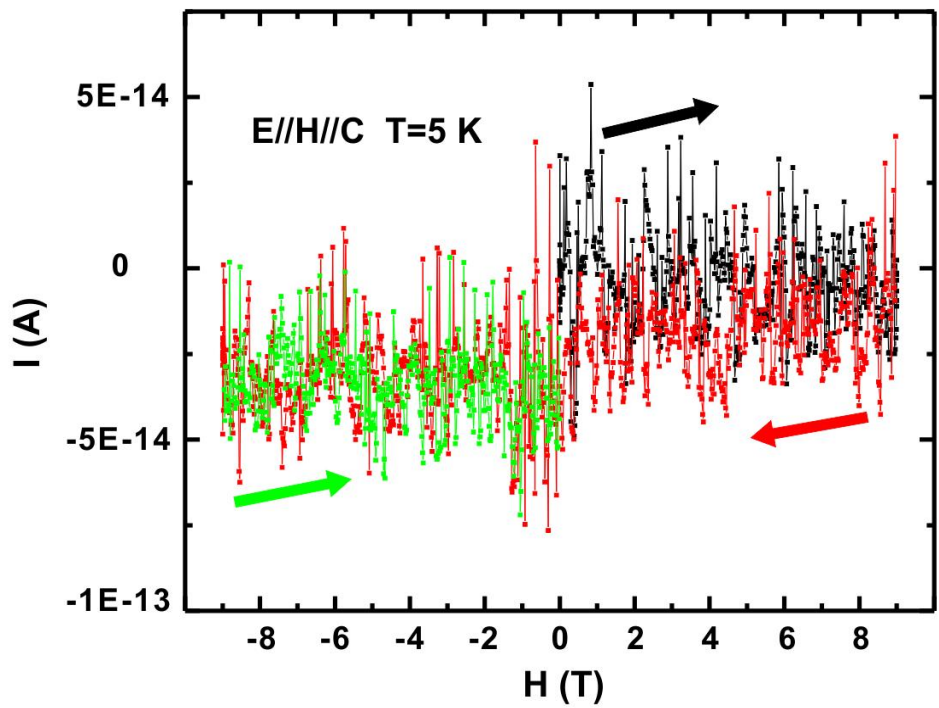

**Figure S8: Current vs. magnetic fields of LSFO2 at 5 K.** E//H//c. The measured current does not respond to the switching of magnetic fields and remains within the noise level. The sequence of the measurement was in the order of black, red, and green arrows.

**Note 3: Estimation of the sensitivity in current measurement with magnetic fields.**

We could estimate the minimum induced current ($I_{min}$) if a 180 $^\circ$ polarization flipping happens when sweeping the field by having the following experimental parameters: electrode contact area (S=0.00021 cm$^2$), sweeping speed of magnetic fields (v=0.02 T/s), the saturated polarization according to PE loops (p=1 μC/cm$^2$). For the worst case, we assume the polarization flips during the whole field sweeping process (from 0 T to 9 T) which takes t= 9 T/ (0.02 T/s) = 450 s. Then $I_{min}$= 2P•S/t = 0.9 pA, which is at least 9 times of our current sensitivity (< 0.1 pA) in the measurement. Therefore, the absence of magnetoelectric current when sweeping magnetic fields is the strong evidence of the decoupling between WFM and FE domains.

**Note 4: Structural topological charge and magnetic topological charge**

A topological vortex concept is essential to understand the decoupling between FE and WFM domains. Here, structural distortions and spins of the identical sites by unit-cell translations (e.g. sites with darker grey triangle backgrounds in Figure 4a and 4b) need to be considered and compared. As the structural (e.g. oxygen) distortion directions (defined by the angle Φ in Figure 4) rotate 60 $^\circ$ between two adjacent domains, the total Φ around the merging point becomes ±2π which can define a topological charge (or the winding number $n_s$) of ±1. The sign of ±2π (or ±1 of the topological charge) depends on the sense of rotation. For example, if vortices that rotate clockwise are defined to have +1 topological charges, then anti-vortices that rotate counter-clockwise have −1 topological charges. The details of these vortex domains as topological objects can be also found in ref. 19 and ref. 23.

In addition, antiferromagnetic spins also rotate around a merging point, similar to what happens with structural distortions. Thus, magnetic domains can also be considered as a topological object and its magnetic topological charge is well defined [2]. However, two spin configurations are possible: in one case, spins rotate by $60°$ between two adjacent domains (i.e. the magnetic topological charge $n_m$ of $\pm1$ around a merging point, Figure 4b), and spins rotate by $120°$ between two adjacent domains in the other case (i.e. the magnetic topological charge $n_m$ of $\pm2$ around a merging point, Figure 4a). Our MFM results are completely consistent with the former even though the theory proposed the possibility of both. This indicates that the magnetic topological charge ($n_m$) tends to be identical with the structural topological charge ($n_s$), which leads to the decoupling of FE and WFM domains. Therefore, we propose it is necessary to consider topological properties when ME coupling is studied in $h$-LuFeO$_3$.